\documentclass[11pt]{article}
\usepackage{moriond,epsfig}

\def\PLB{{\em Phys. Lett.}  B}

\def\PRD{{\em Phys. Rev.} D}

\def\be{\begin{equation}}
\def\ee{\end{equation}}
\def\bea{\begin{eqnarray}}
\def\eea{\end{eqnarray}}

\begin{document}
\vspace*{4cm}
\title{The first polarized Proton Collisions at the STAR experiment at RHIC}

\author{Bernd Surrow\footnote{surrow@bnl.gov}}

\address{Department of Physics, Brookhaven National Laboratory\\
Upton, NY 11973-5000, USA}

\maketitle\abstracts{
 The first run of transverse polarized protons at RHIC was recently completed which opened
 a new era exploring the spin structure of the proton.
 A first measurement of the single transverse spin asymmetry, $A_{N}$, for leading $\pi^{0}$
 production
 from transverse colliding polarized protons at $\sqrt{s}=200\,$GeV, $x_{F}>0.25$ and $p_{T}\simeq 1-4\,$GeV
 was a focus of the STAR collaboration during the first polarized proton run at RHIC.  	
 Two new subcomponents have been added to the STAR experiment to carry out such a measurement
 in polarized proton collisions:
 a forward $\pi^{0}$ detector system at approximately $7.8\,$m of the STAR interaction region
 to reconstruct $\pi^{0}$ mesons from their decay products ($\pi^{0} \rightarrow \gamma \gamma $) and
 a beam-beam counter with large forward acceptance to provide a means of beam-related background suppression
 and relative luminosity measurement.
}

\section{Introduction}
The spin of elementary particles is as fundamental to their nature as their mass. 
The proton is a fermion of $J=1/2$. The proton itself consists of valence
quarks, gluons and quark-antiquark pairs, known as the sea. Quarks and gluons are the fundamental
ingredients of QCD.
Unpolarized electron-proton collider experiments (ZEUS/H1) and 
several lepton-nucleon fixed-target experiments 
have played an important role in our current understanding of hadronic matter. 
Similarly to the unpolarized case, several polarized fixed-target experiments have been
conducted in the past to gain a deeper understanding of the spin structure of the proton. 
Those experimental efforts have been restricted to large values of Bjorken $x$. The proton
spin is understood to be made up of contributions arising from the quark spin, the
gluon spin and orbital angular momentum.
The fundamental question in this regard is how the proton spin is distributed among those 
contributions. It was found in polarized lepton-nucleon experiments 
that only about $1/3$ of the proton spin is carried by quarks and anti-quarks,
contrary
to the expectation of the constituent quark model
that the proton spin would be carried dominantly by its three valence quarks.
A significant fraction of the proton spin must therefore be carried by gluons and orbital angular momentum.
The role of the gluons to make up for the missing proton spin is currently only very poorly constrained
from scaling violations in fixed target experiments.
A need for a new generation of experiments to explore the spin structure of the proton is clearly 
apparent. The current spin physics effort at RHIC at BNL focuses on the collision of polarized protons
to gain a deeper understanding of the spin structure of the proton in a new, previously unexplored
territory.
The first polarized proton run from December 2001 until January 2002 is the beginning of a multi-year
experimental program which aims to address a variety of topics related to the spin structure of the 
proton such as: 1. spin structure of the proton (gluon contribution of the proton spin, 
flavor decomposition of the quark and anti-quark polarization and transversity distributions of the proton),
2. spin dependence of fundamental interactions, 3. spin dependence of fragmentation and 
4. spin dependence of elastic polarized proton collisions.
A recent review of the RHIC spin program can be found in \cite{bunce}.

The principle approach to study spin effects is to measure an asymmetry (A) which quantifies
the normalized difference of measured yields for different initial-state spin configurations. Ultimately, any combination
of beam polarization, i.e. either longitudinal (L) or transverse (T), will be possible at RHIC
to access different aspects of the proton spin structure. 
A crucial fact to remember is that the statistical significance of double spin asymmetries varies as $P^{4}\int Ldt$ whereas
for single spin asymmetries it varies as $P^{2}\int Ldt$. Thus, the demand on high polarization is particularly important
for the measurement of double spin asymmetries.
In the following, two prominent examples of asymmetry measurements in polarized proton collisions will 
be discussed.

The measurement of the double longitudinal spin asymmetry, $A_{LL}$, for photon production allows the extraction of the gluon 
polarization, $\Delta G/G$. $A_{LL}$. In LO QCD, employing factorization of the underlying
hard process the asymmetry measured for $\vec{p}+\vec{p} \rightarrow \gamma + {\rm jet} + X$ is represented as: 
$A_{LL}=\frac{\Delta G(x_{g})}{G(x_{g})} \cdot A_{1}^{p}(x_{q})\cdot \hat{a}(g+q \rightarrow \gamma+q)$.
The ratio of the polarized and unpolarized structure function, $A_{1}^{p}(x_{q})$,
is measured in polarized deep-inelastic scattering and 
$\hat{a}(g+q \rightarrow \gamma+q)$ is calculated in pQCD.
Hence $A_{LL}$ for prompt photons detected in coincidence with the away-side quark-jet allows an extraction of the 
gluon polarization $\Delta G(x_{g})/G(x_{g})$. 

A focus of the first polarized proton run at the STAR experiment was the measurement of a single transverse spin asymmetry, $A_{N}$,
to achieve a first polarization observable measurement at RHIC.
Non-zero values for $A_{N}$ have been observed at the FNAL E704 \cite{E704} experiment for $\vec{p} + p \rightarrow \pi + X$
at $\sqrt{s}=20\,$GeV and $0.5<p_{T}<2.0\,$GeV. Theoretical models that explain the E704 data
also predict non-zero values for $A_{N}$ for pion production at RHIC.
Qiu and Sterman \cite{qiu_sterman}
attribute the measured asymmetry to a higher-twist pQCD effect. The group of Anselmino and
Leader \cite{anselmino} perform a global analysis of semi-inclusive DIS data from HERMES \cite{HERMES} and E704 data. 
This approach involves initial
(`Sivers effect') as well final (`Collins effect') state fragmentation effects to account as possible explanations
for the measured asymmetries.

Besides the theoretical interest in measuring $A_{N}$, it could serve as a potential candidate to monitor the RHIC 
beam polarization at a particular experiment (`local polarimeter').

\begin{figure}[t]
\setlength{\unitlength}{\textwidth}
\begin{picture} (0.7,0.2)
\put (0.05,0.0){\mbox{\epsfig{figure=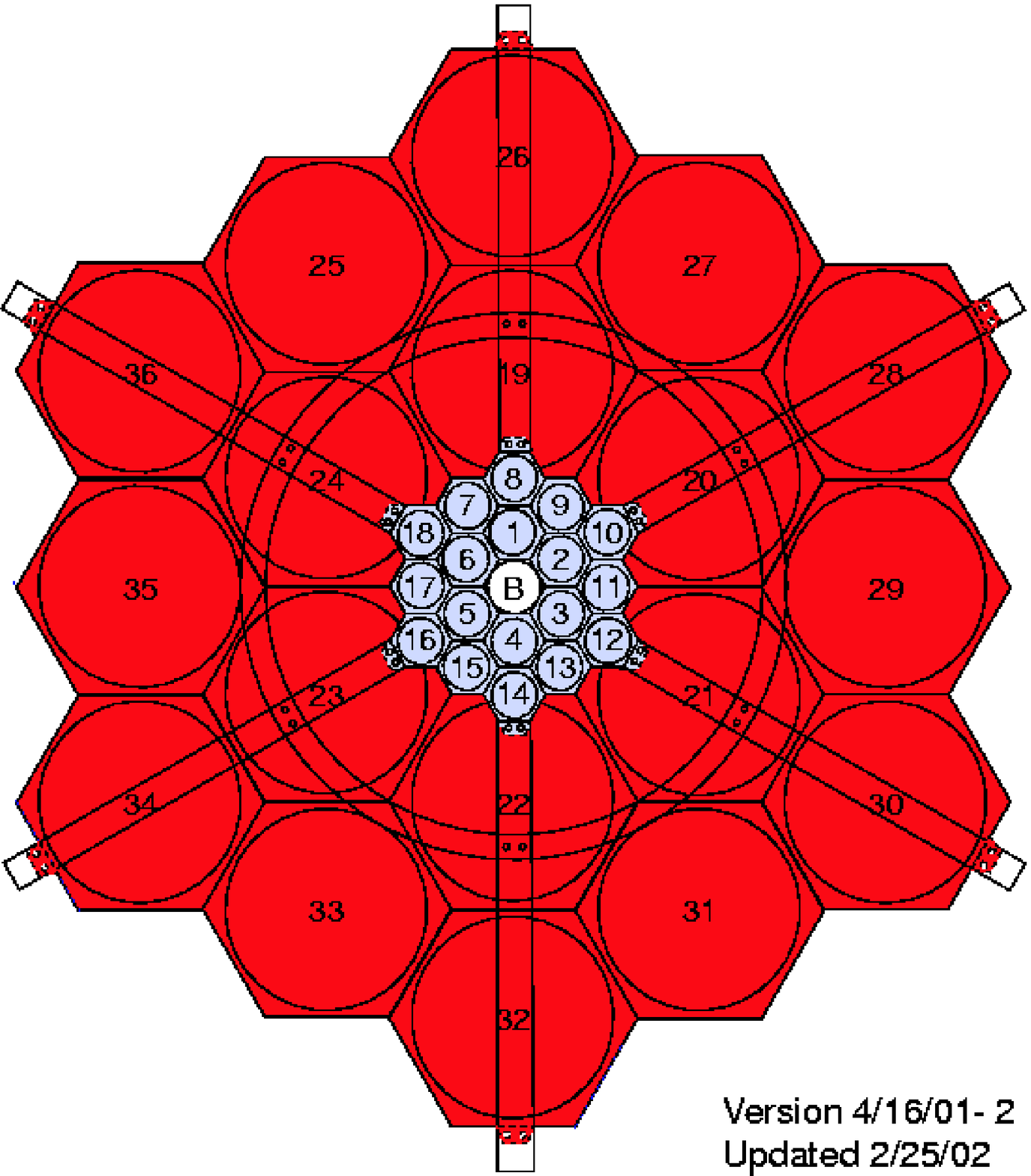,width=3.5cm,clip=}}}
\put (0.60,0.0){\mbox{\epsfig{figure=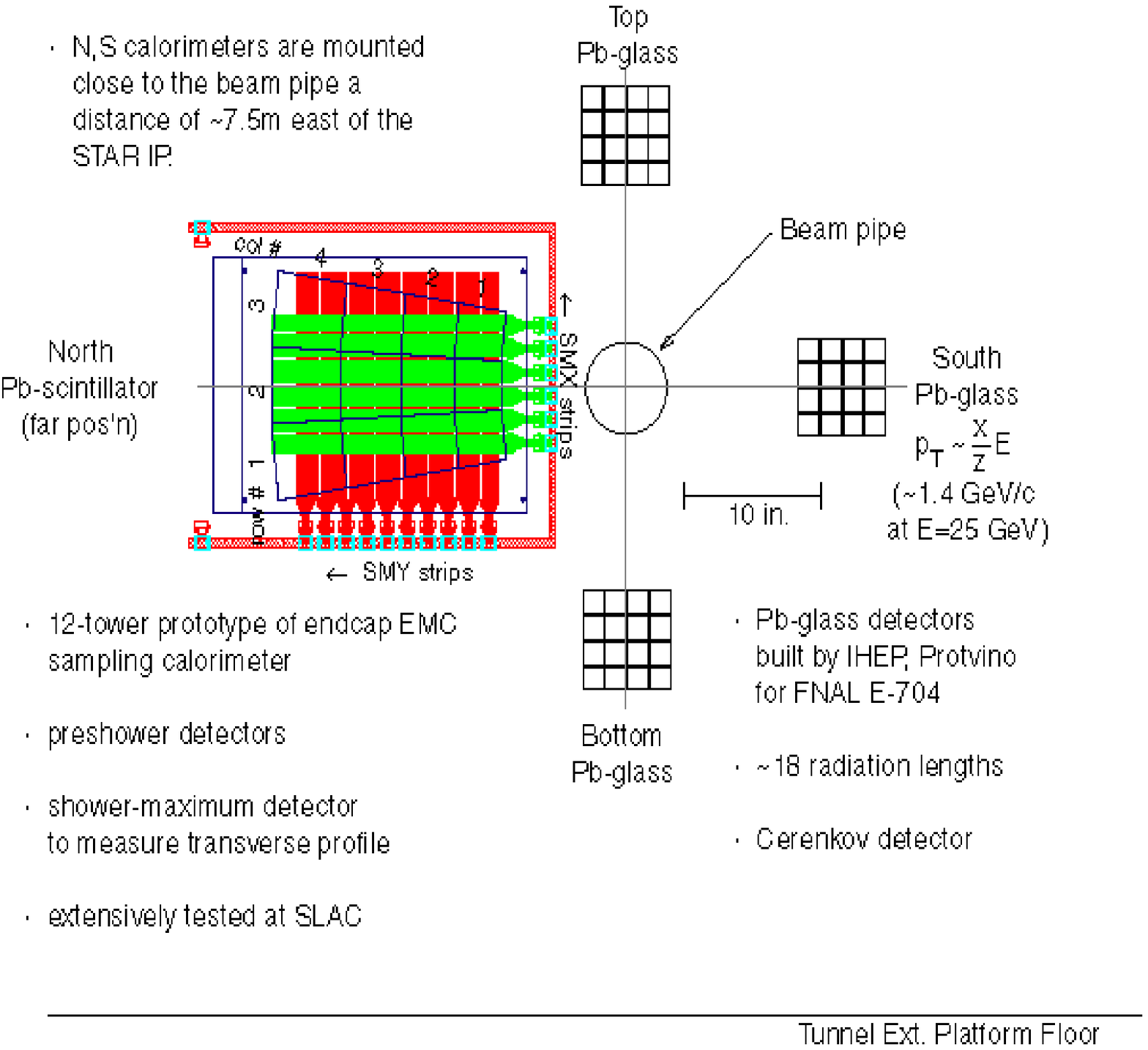,width=4.0cm,clip=}}}
\end{picture}
\caption{\it Schematic of the STAR beam-beam counter BBC (left) and forward pion detector (FPD) (right).}
\label{fpd_bbc_sketch} 
\end{figure}

\section{The polarized proton collider RHIC}

The first collisions of polarized protons occurred in December 2001, ushering in a 
a new era to complement the ongoing relativistic heavy-ion program. RHIC is the first accelerator
to accelerate and collide polarized protons, ultimately at high luminosity, at a center-of-mass energy
of up to $500\,$GeV. 

The key to maintain the proton polarization through acceleration despite its large anomalous magnetic
momentum, is to perform a rotation of the proton spin by $180^{\circ}$ in the horizontal plane around a particular axis.
This manipulation is performed by helical dipole magnets, known as `Siberian snakes', which have been used for the first time
at a proton collider. With two Siberian snakes installed in each ring, cumulative tilt effects of the proton spin are canceled,
thereby eliminating the influence
of depolarizing spin resonances. Besides the installation of Siberian snakes, the PHENIX
and STAR experiments will be equipped with spin rotator magnets to allow for the precession from transverse to longitudinal
polarization and thus to collide longitudinal polarized proton beams.

The first polarized proton run at RHIC was carried out at a center-of-mass energy of $200\,$GeV.
Each ring was loaded with 55 bunches of alternating polarization resulting in a 
a bunch crossing-time of $214\,$ns. A transverse polarization 
of about $20\,\%$ was achieved at the injection energy of $24.6\,$GeV and was approximatley maintained when the protom 
beams were accelerated to $100\,$GeV. 

\section{Upgrade of the STAR detector for the first polarized proton run}

The goal of the first RHIC polarized proton run at the STAR experiment was 
the measurement of the single-transverse spin asymmetry $A_{N}$ for forward $\pi^{0}$ production
at $x_{F} \simeq 0.1 - 0.6$ and $p_{T} \simeq 1 - 4\,$GeV. $A_{N}$ is extracted from :
\begin{equation}
A_{N}=\frac{1}{P}\frac{N^{\rm \uparrow}-R\cdot N^{\rm \downarrow}}{N^{\rm \uparrow}+R\cdot N^{\rm \downarrow}}
\end{equation}
which requires three independent measurements:
1. the spin-dependent yields ($N^{\rm \uparrow (\downarrow)}$) of forward $\pi^{0}$ production, 2. the relative luminosity
$R=L^{\rm \uparrow}/L^{\rm \downarrow}$ and 3. the actual beam polarization $P$. 

The latter is the focus of a dedicated effort at RHIC to obtain a fast (relative) polarization measurement using 
pC elastic scattering at very small $|t|$ values.
This Coulomb Nuclear Interference polarimeter will ultimately be calibrated
to $pp$ elastic scattering for a polarized hydrogen gas-jet target.

An upgrade program at the STAR experiment was performed with the installation of a beam-beam counter (BBC) and
a forward-pion detector (FPD). As well, the commissioning of the barrel electromagnetic calorimeter (BEMC) modules and trigger was
undertaken. Ultimately, the BEMC will play a crucial role in measuring prompt photons and jets.
In addition, a spin scaler system was commissioned to account for the beam polarization reversals every
bunch crossing of $214\,$ns. 

A layout of the BBC can be seen in Figure \ref{fpd_bbc_sketch} (left). It consists of a hexagonal scintillator array structure at
$\pm 3.5\,$m from the nominal interaction point. The BBC is the main device to make the relative luminosity measurement
and to provide a trigger to distinguish $\vec{p}\,\vec{p}$ events from beam related background events by means of timing 
measurements.

The FPD system, shown in Figure \ref{fpd_bbc_sketch} (right), consists of three lead-glass electromagnetic calorimeter modules 
together with a lead-scintillator 
calorimeter, which is a prototype module of the STAR endcap calorimeter. The latter device 
allows the reconstruction of $\pi^{0}$ mesons from their decay products ($\pi^{0} \rightarrow 
\gamma \gamma $) by measuring the total energy and the transverse shower profile.
It consists of 12 independent towers, pre-shower detectors and a shower-maximum detector 
to perform a transverse
shower profile measurement. This prototype module has been extensively 
studied using high energy electron test beams at SLAC. Its testbeam
performance is well reproduced by a GEANT simulation.

\begin{figure}[t]
\setlength{\unitlength}{\textwidth}
\begin{picture} (0.7,0.25)
\put (0.1,0.0){\mbox{\epsfig{figure=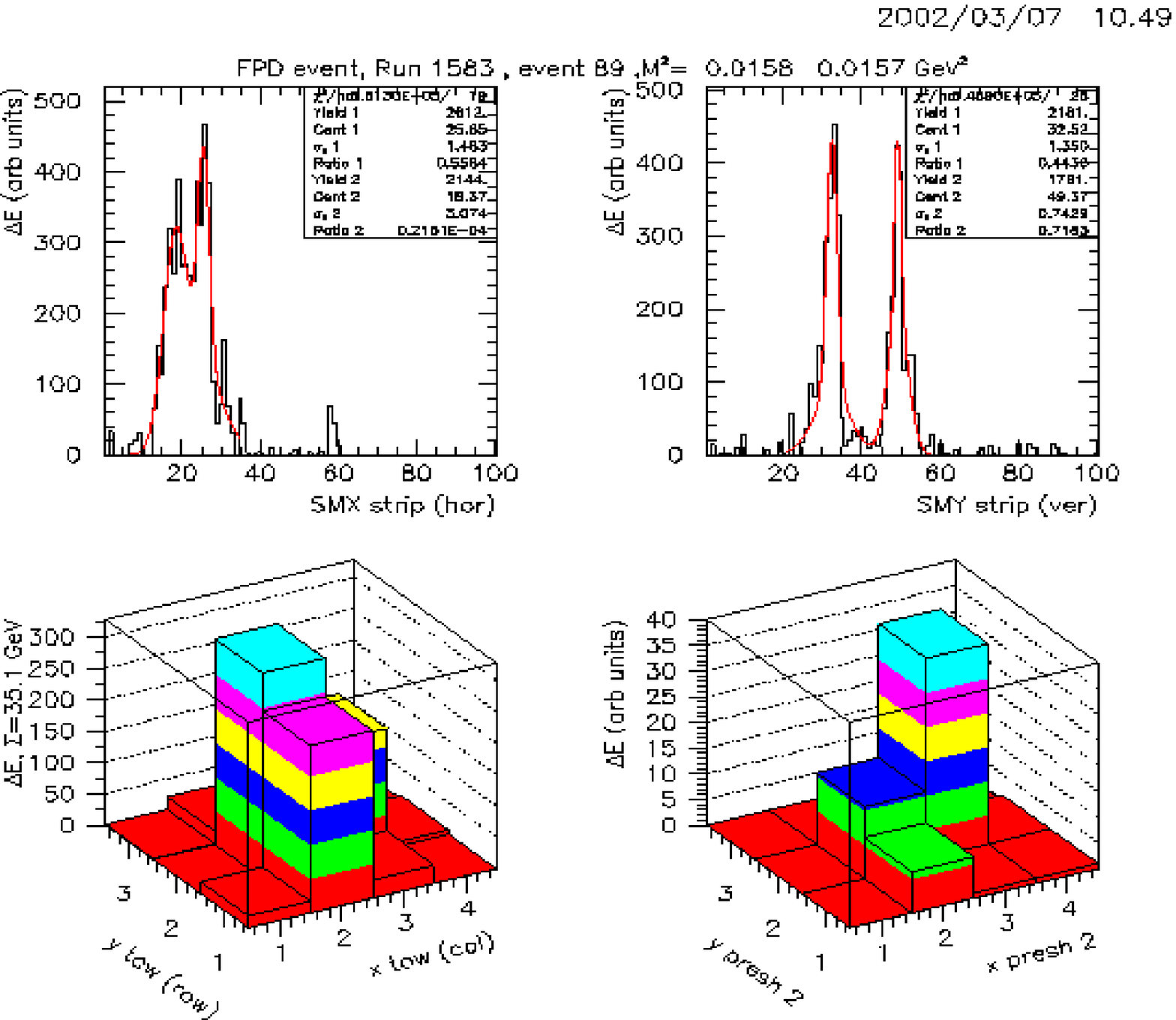,width=5.0cm,clip=}}}
\put (0.55,0.0){\mbox{\epsfig{figure=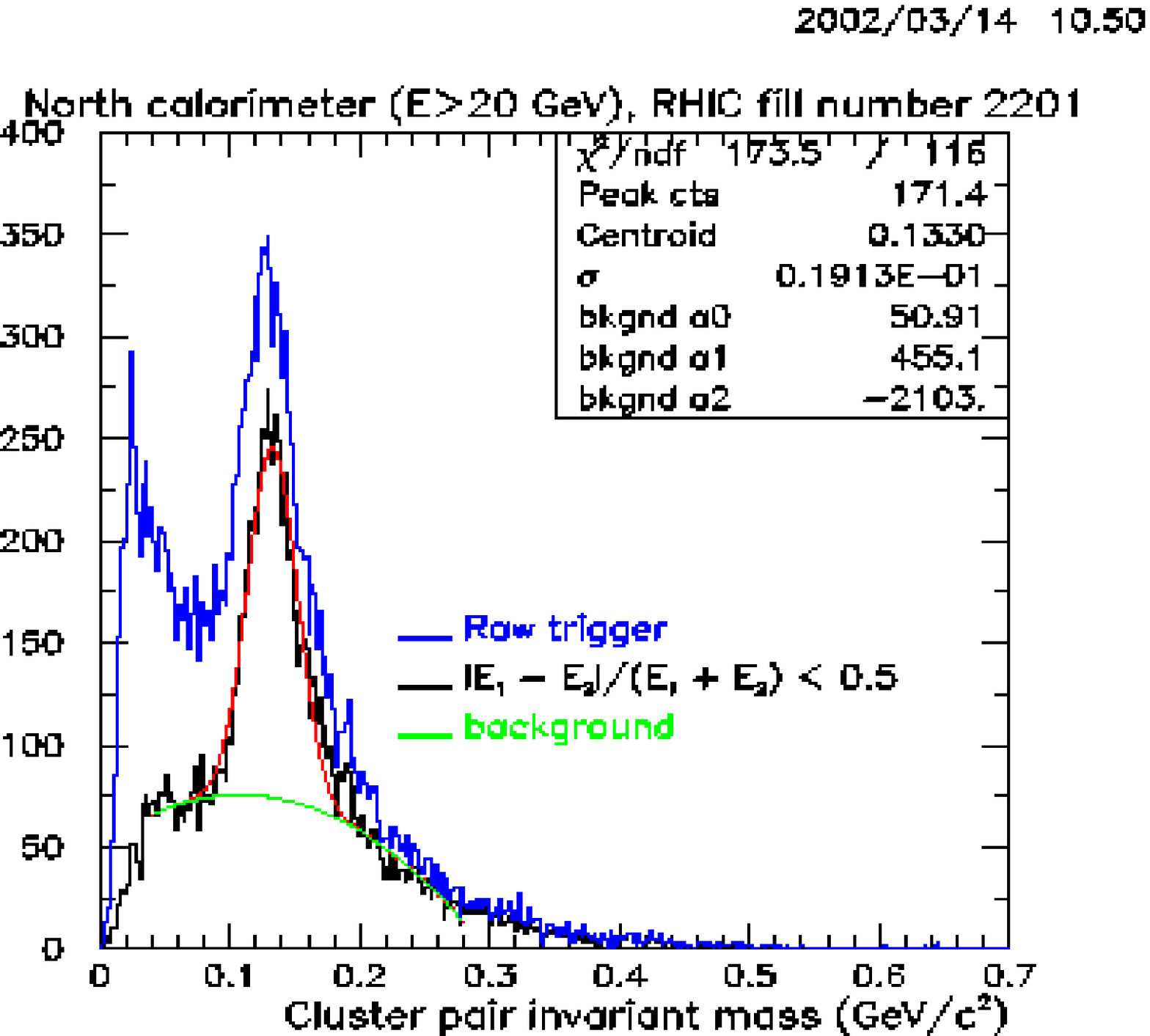,width=5.0cm,clip=}}}
\end{picture}
\caption{\it FPD transverse shower profile (left) and invariant mass distribution of $\pi^{0}$ 
candidate events (right).}
\label{fpd_perf} 
\end{figure}

The FPD transverse shower profile for a typical event of the shower maximum detector is shown in 
Figure \ref{fpd_perf} (left) together
with the calorimeter and pre-shower detector response. The cluster separation measured in the shower maximum detector and 
the measured
calorimeter energy serves as input to the actual $\pi^{0}$ mass determination. $\pi^{0}$ mesons of up to $60\,$GeV 
have been reconstructed. 
A clearly identified $\pi^{0}$ mass peak can 
be seen in Figure \ref{fpd_perf} (right). Those results are very encouraging to extract $A_{N}$ from forward $\pi^{0}$ 
production
and thus to study expected spin dependent effects in polarized proton collisions at RHIC. 

\section{Summary and outlook}

The first polarized proton run at RHIC started a new era at BNL of exploring the spin structure
of the proton. The main focus of the STAR experiment during the first polarized proton run of transverse polarization 
was the measurement of a single transverse spin asymmetry 
of forward $\pi^{0}$ production. In preparation of the first polarized proton run, 
an upgrade of the STAR experiment was performed with the installation of a beam-beam counter and a forward
pion detector system, besides the commissioning of a spin scaler system. 
First results show a clear identification of forward produced $\pi^{0}$ mesons.

The STAR detector will undergo major upgrade programs with the installation of the endcap
calorimeter which is the principal device to explore the gluon polarization of the proton and the
barrel calorimeter besides a completion of the beam-beam counter. An upgrade proposal for a new
forward pion detector system is currently under preparation. 

An exciting time is ahead of us to explore the spin structure of the proton in a previously
unexplored territory.


\end{document}